\documentclass[preprint,showpacs,showkeys,aps,prd,amsfonts,eqsecnum,nofootinbib,superscriptaddress]
{revtex4} 
\usepackage{graphicx,epsfig}
\newcommand{\Rslash}{R\kern -6.4pt\big{/}}

\begin{document}
%
\preprint{CUMQ/HEP 133, WM-05-107}
%
%
\title{Neutrino Masses in the Effective Rank-5 Subgroups of $E_6$ II: Supersymmetric Case}
\author{Mariana Frank}
\email[]{mfrank@vax2.concordia.ca}
\affiliation{Department of Physics, Concordia University, 7141 Sherbrooke St. 
West, Montreal, Quebec, CANADA H4B 1R6}
\author{Marc Sher}
\email[]{sher@physics.wm.edu}
\affiliation{Particle Theory Group, Department of Physics,
College of William and Mary, Williamsburg, VA 23187-8795, USA}
\author{Ismail Turan}
\email[]{ituran@physics.concordia.ca}
\affiliation{Department of Physics, Concordia University, 7141 Sherbrooke St.
West, Montreal, Quebec, CANADA H4B 1R6}
\date{\today}

\begin{abstract}
We present a complete analysis of the neutral fermion sector of supersymmetric $E_6$-inspired low 
energy models containing an extra $SU(2)$, concentrating on the Alternate Left-Right and Inert models.  
We show that the R-parity conserving scenario always exhibits a large Dirac mass for $\nu_L$ with 
maximal mixing with an isosinglet neutrino, and that R-parity violating scenarios do not change the 
picture other than allowing further mixing with another isosinglet.   In order to recover Standard Model 
phenomenology, additional assumptions in the form of discrete symmetries and/or new interactions 
are needed.   We introduce and investigate Discrete Symmetry method and Higher Dimensional 
Operators as mechanisms for solving the neutrino mass and mixing problems in these models.\end{abstract}
\vskip -2.5cm
\maketitle
\section{Introduction}\label{sec:intro}

Supersymmetric grand unified theories are among the most attractive 
scenarios for physics beyond the Standard Model.  They are well motivated by 
superstring theories which may lead to a consistent theory of all 
interactions.
Of these, SU(5) models have been studied extensively. The minimal 
SU(5) models predicted too large a decay rate for the proton and had to 
be modified. More recently, doubts have been raised about the 
validity of even modified SU(5) models, due to 
the
discovery of solar \cite{solar} and atmospheric \cite{atm} 
neutrino oscillations. Small neutrino masses can be explained most 
elegantly through the seesaw mechanism, which requires the presence of 
a right-handed neutrino, a particle not naturally present in the 
spectrum of SU(5).
Though scenarios with an extended neutrino sector 
exist in SU(5), it is worthwhile investigating Grand Unified Theories (GUTs) which naturally
contain the right-handed neutrino. Experimental data from 
the Los Alamos Liquid
Scintillation Detector (LSND) requires neutrino mass square splittings 
\cite{LSND} which are 
in serious disagreement with other results unless one or more 
neutrinos are added and are ``sterile" \cite{sterile}.
Such scenarios have been
studied extensively \cite{Sorel:2003hf, Mohapatra:2004uy, 
Babu:2004mj, Godfrey:2004gv, Krolikowski:2004ru, McDonald:2004pa,
Barger:2003xm, Stephenson:2003ta, Paes:2002ah, Donini:2001qv}. Since the 
mixing of sterile and active neutrinos affects the interpretation of 
results from solar and atmospheric neutrino experiments,  limits have 
been set on
such mixings.   

Sterile neutrinos can occur naturally in 
supersymmetric GUTs, which often
  predict the existence of exotic
fermions. Of these, superstring-inspired $E_6$ is one of the most 
attractive choices. $E_6$ is the next anomaly-free choice group
after $SO(10)$.  It is based on an exceptional Lie group with complex 
representations, where each generation of fermions can be placed in
the {\bf 27}-plet representation. 

The $E_6$ spectrum contains several
neutral exotic fermions, some which could be interpreted as sterile
neutrinos. The precise details of mass generation and mixing with the
active neutrinos would depend the particular subgroup of $E_6$
considered.  There are many phenomenologically acceptable low energy
models which arise from $E_6$. In this work we concentrate on rank-5
subgroups, which always break to $SU(3)_c \times SU(2)_L \times
U(1)_Y \times U(1)_{\eta}$ \cite{HR,Langacker}, and which contain an 
extra $SU(2)$ symmetry in addition to the MSSM symmetry.
  These intermediate subgroups  give rise to the usual Supersymmetric 
Left-Right Model (LRSUSY) \cite{LRSUSY}, the
Alternative Left-Right Supersymmetric model (ALR) \cite{ALR} and the 
Supersymmetric Inert model \cite{inert}.  In a previous work we have 
shown that, contrary to expectation, and despite a rich exotic sector 
in the neutral fermionic sector containing three extra states, the 
non-supersymmetric version of these models did not provide neutrino 
masses and mixing consistent with neutrino experiments 
\cite{Frank:2004vg}.  In this paper we analyze  masses and mixings of 
neutrinos in supersymmetric $E_6$ inspired models. Before we proceed, 
we summarize our previous results.

In the non-supersymmetric version of either the ALR or the 
Inert model (and the discussion is the same for the LR model), the 
lightest state in the neutral fermion sector contains only $SU(2)_L$ 
singlets, which do not interact with SM particles. Additionally, the 
models predict two more light neutrino states with masses of the 
order of the up quark mass. These are phenomenologically unacceptable. 
In order to cure these problems, 
additional symmetries and/or new interactions are needed. In the 
simplest such non-minimal scenario, the Discrete Symmetry method requires 
imposing one extra discrete symmetry only. The aim is to eliminate the 
tree-level Dirac mass in the Lagrangian, thus generating radiative 
masses only for neutrinos. This method requires an extra $SU(2)_L$ 
Higgs doublet with vanishing vacuum expectation value (vev). It cures the Dirac neutrino mass problem, 
but predicts large mixing between active and sterile states.

The second 
method, the Higher Dimensional Operators (HDO) method, requires
additional Higgs fields from the ${\overline{\bf{27}}}$-plet of $E_6$ and the
existence of some intermediate scale. Higher dimensional 
(dimension-5) operators induce interactions which are suppressed by
  one power of the compactification scale.
  This method solves  the neutrino mass problems but does
not predict any sterile component(s) in the lightest neutrino
state, which is now an admixture of $\nu_L$ and $N_L$, an exotic 
($SU(2)_L$ doublet) particle. The effect of this mixing is
  to lower the electron neutrino coupling to the electron and the $W_L$. 
There exist similar reductions for the muon and tau couplings. Furthermore, 
when the reduction is different for each generation, this will violate lepton 
universality. See Ref. \cite{Frank:2004vg} for details.

The last method introduced is the Additional Neutral Fermion
(ANF), which requires the existence of both new particles and new discrete
symmetries. The additional
interactions  are of the type
${\bf{27}}\cdot {\bf{\overline{27}}}\cdot {\bf{1}}$,  which further
require additional Higgs doublets from the
${\bf{27}}+{\bf{\overline{27}}}$ representation. In order not to
alter existing couplings, the vev's of the new fields must
be chosen suitably, and an additional $Z_2$ symmetry is needed. In
these circumstances,  we obtain  two light states with an active neutrino
part of the form predicted by the HDO method, but
mixed
with a sterile flavor state. The
mixing is completely arbitrary.
Extended to three generations, the model contains two
structures, ${\bf{2+2}}$ and ${\bf{3+1}}$, or, if
the above mixing is sizable only for one generation, the
${\bf{2+2}}$ structure
arises naturally. Otherwise,
more
realistically, including three generations for each exotic neutral 
fermion, we obtain a ${\bf{3+3}}$ structure.

  This 
paper is organized as follows. We discuss supersymmetrized versions of the 
Alternative Left-Right  and Inert models in Section~\ref{sec:model}. 
In Section~\ref{sec:ALRI} we analyze neutrino masses and mixings in the 
ALR and Inert models within the R-parity conserving scenario. Both of these 
models suffer from predicting too large  a Dirac mass for the active 
neutrinos. The possible mixing between $R=+1$ and $R=-1$ sectors through
soft R-parity violating terms is discussed for each models separately in Section~\ref{sec:Rviol} and in an Appendix.
All possible hierarchies among the parameters exhibit the feature that 
the physically relevant state still has too large a mass and the lightest state is fully sterile.
So, in Section~\ref{sec:mixing}, we suggest mechanisms for going beyond the minimal content of the models 
in order to rectify this problem. We conclude and summarize our results in Section~\ref{sec:conc}.

\section{Description of the Models}\label{sec:model}

The details of the models are given in our earlier work \cite{Frank:2004vg}. Here we would like  to summarize our previous results and concentrate on  the Higgs sectors of the SUSY models where the difference occurs with respect to their non-SUSY versions.

Under the maximal subgroup $SU(3)_C\otimes SU(3)_L\otimes SU(3)_H$ of $E_6$, 
the $\bf{27}$ dimensional representation of $E_6$ branches into 
\begin{eqnarray}
\bf{27}&=&({\bf 3^c},{\bf 3},{\bf 1})+({\bf\bar{3}^c},{\bf 1},{\bf\bar{3}})+({\bf 1^c},{\bf\bar{3}},{\bf 3})\nonumber\\
&=&\;\;\;\;\;\;q\;\;\;\;\;+\;\;\;\;\; \bar{q}\;\;\;\;\;\;+\;\;\;\;\;l\,,
\label{333}
\end{eqnarray}
where
\begin{eqnarray}
q=\left(\begin{array}{c}
u \\
d \\
h
\end{array}
\right)_L,\;\;
\bar{q}=\left(u^c\;d^c\;h^c\right)_L,\;\;
l=\left(\begin{array}{ccc}
E^c & N & \nu \\
N^c & E   & e \\
e^c & \nu^c & S^c
\end{array}\right)_L.
\end{eqnarray}
Here $SU(3)_H$ operates horizontally. There are three different ways to break $SU(3)_H$
into  $SU(2)_H\otimes U(1)_{Y_H}$. When the first and the second columns form a $SU(2)_H$ doublet, the so-called
Left-Right (LR) symmetric model is obtained. Its alternative version is when the first and the third columns form a doublet, which is the Alternative Left-Right (ALR) symmetric model. The last combination is when the second and the 
third  columns combine to form a doublet and the Inert model is obtained. See Ref. \cite{Frank:2004vg} for more details. 

In the LR and ALR models, both $SU(2)_H$ and $U(1)_{Y_H}$ contribute to the electromagnetic charge $Q_{em}$. In the Inert model, however, $SU(2)_H$ does not contribute  to $Q_{em}$, which leads to neutral gauge bosons and a very different phenomenology \cite{inert}. We will use the notation $H=R,R^{\prime},I;
Y_H=Y_{R,R^{\prime},I}$ for the LR, ALR and Inert groups, respectively. We consider their rank-5 versions whose gauge groups 
are $SU(3)_C\otimes SU(2)_L\otimes
SU(2)_R\otimes U(1)_V,\,SU(3)_C\otimes SU(2)_L\otimes
SU(2)_{R^{\prime}}\otimes U(1)_V$, and
$SU(3)_C\otimes SU(2)_L\otimes SU(2)_I\otimes U(1)_Y$
for LR, ALR and Inert cases, respectively. The quantum numbers of the particles in the ALR and Inert models are given in Table \ref{tablepartic}. 
\begin{table}[t]
	\caption{The quantum numbers of fermions in $\bf{27}$ of $E_6$ at $SU(3)_C\otimes SU(2)_L\otimes SU(2)_{R^{\prime}}\otimes U(1)_{V=Y_L+Y_{R^{\prime}}}$ and $SU(3)_C\otimes SU(2)_L\otimes SU(2)_I\otimes U(1)_Y$ levels.} \label{tablepartic}     
\begin{center}
    \begin{tabular}{ccccccc}
    \hline\hline
\vspace{0.1cm}
$\;\;\;\;\;$ state $\;\;\;\;\;$ & $\;\;\;\;\;I_{3L}\;\;\;\;\;$ & $\;\;\;\;\;I_{3R^{\prime}}\;\;\;\;\;$ & $\;\;\;\;\;I_{3I}\;\;\;\;\;$ & 
$\;\;\;\;\;V/2\;\;\;\;\;$ & 
$\;\;\;\;\;Y/2\;\;\;\;\;$ & $\;\;\;\;\;Q_{em}\;\;\;\;\;$\\ 
\hline 
    $u_L$      &  1/2    &  0     &  0     &  1/6   & 1/6   &  2/3     \\
    $u_L^c$    &  0      &  -1/2   &  0     &  -1/6  & -2/3  &  -2/3     \\
    $d_L$      &  -1/2   &  0     &  0     &  1/6   &  1/6  &  -1/3     \\
    $d_L^c$    &  0      &  0     &  -1/2   &  1/3   &  1/3  &  1/3    \\
    $h_L$      &  0      &  0     &  0     &  -1/3  &  -1/3 &  -1/3     \\
    $h_L^c$    &  0      &  1/2  &  1/2  &  -1/6   &  1/3  &  1/3     \\
    $e_L$      &  -1/2   &  -1/2  &  -1/2  &  0     &  -1/2 &  -1     \\
    $e_L^c$    &  0      &  1/2   &  0     &  1/2   &  1    &  1     \\
    $E_L$      &  -1/2   &  0     &  1/2   &  -1/2  &  -1/2 &  -1     \\
    $E_L^c$    &  1/2    &  1/2   &  0     &  0     &  1/2  &  1     \\
    $\nu_L$    &  1/2    &  -1/2  &  -1/2  &  0     &  -1/2 &  0     \\
    $\nu_L^c$  &  0      &  0     &  1/2   &  0     &  0    &  0     \\
    $N_L$      &  1/2    &  0     &  1/2   &  -1/2  &  -1/2 &  0     \\
    $N_L^c$    &  -1/2   &  1/2   &  0     &  0     &  1/2  &  0     \\
    $S_L^c$    &  0      &  -1/2  &  -1/2  &  1/2   &  0    &  0     \\
\hline \hline
        \end{tabular}
        \end{center}
\vskip -0.5cm
\end{table}

The Higgs sector of $E_6$ in the SUSY scenario differs from the non-SUSY case. Since SUSY requires doubling the 
number of particles, there exist many scalar fields, some of which may be taken as the Higgs bosons required for symmetry breaking. In fact, there are two ways to proceed \cite{HR}. One could assign the Higgs fields to the same ${\bf{27}}$ (or to a $\bf{\overline{27}}$) as the usual fermions and then some of the superpartners of the fermions can play the role of the Higgs fields. Or, it is possible to assign them to  different  ${\bf{27}}$ representations than the fermions, and the Higgs fields are introduced as additional scalars. The latter is less economical than the former and very similar to the non-SUSY case which was discussed in the earlier paper \cite{Frank:2004vg}. So we choose to work in the former framework. In fact, our approach is to choose as many Higgs bosons as possible among the superpartners of lepton fields and consider other scalars from different ${\bf 27}$'s only if necessary.

To analyze the Higgs sector further, we need to write the most general $R$-parity conserving renormalizable superpotential invariant under the Standard Model gauge group \cite{HR} 
\begin{eqnarray}
W&=&W_1+W_2+W_3+W_4\,,\nonumber\\
W_1&=&\lambda_1 Q u_L^c H^c +\lambda_2 Q d_L^c H+\lambda_3 L e_L^c H+\lambda_4 HH^cS_L^c +\lambda_5 h_L h_L^c S_L^c\,,\nonumber\\
W_2&=&\lambda_6 h_L u_L^c e_L^c +\lambda_7  L Q h_L^c +\lambda_8 d_L^c \nu_L^c  h_L \,,\nonumber\\
W_3&=&\lambda_9 QQh_L +\lambda_{10}h_L^cu_L^cd_L^c\,,\nonumber\\
W_4&=&\lambda_{11}LH^c\nu_L^c\,,\label{SP}
\end{eqnarray}
where the following notation is used:
\begin{eqnarray}
Q&=&\left(\begin{array}{c}u\\d\end{array}\right)_L \,,\;L=\left(\begin{array}{c}\nu\\e\end{array}\right)_L\,,\:H=\left(\begin{array}{c}N\\E\end{array}\right)_L\,,\;
H^c=\left(\begin{array}{c}E^c\\N^c\end{array}\right)_L.
\end{eqnarray}
In each term in $W$, one of three fields corresponds to a scalar field and thus each term represents three different Yukawa interactions. We later discuss the ALR and Inert models by further imposing $SU(2)$ symmetries on the superpotential $W$, which reduces the number of independent Yukawa couplings. Now, before choosing the Higgs fields to be superpartners of the fermions, we first must determine the baryon ($B$) and lepton ($L$) numbers (and $R$ parity) of the exotic fields (the ones in Table \ref{tablepartic} other than the Standard Model fields). To get consistent $B$ and $L$ assignments, not all of the terms in $W$ can exist simultaneously. Possibilities can be classified with respect to the $(B,L,R)$ assignments of the fields $h_L$ and $\nu_L^c$. The existence of the $W_2$-term requires $(B,L,R)_{h_L}=(1/3,1,-1)$ ($h_L$ is a leptoquark),  the $W_3$-term requires $(B,L,R)_{h_L}=(-2/3,0,-1)$ ($h_L$ is a diquark). Clearly, both the $W_2$ and $W_3$ terms can not exist simultaneously without violating baryon and lepton numbers. If one wished to treat $h_L$ as an ordinary quark\footnote{In fact, no direct constraint comes from the $W_1$-term, but all other considerations lead to a stable $h_L$ which is  phenomenologically problematic. See Ref. \cite{HR} for details.}, $(B,L,R)_{h_L}=(1/3,0,+1)$, then both $W_2$ and $W_3$ would be eliminated. For the $\nu_L^c$ field there are two possibilities; $(B,L,R)_{\nu_L^c}=(0,1,+1)$ or  $(B,L,R)_{\nu_L^c}=(0,0,-1)$. Unlike the former, the latter assignment allows a non-zero vev for the superpartner of $\nu_L^c$, $\tilde{\nu}_L^c$, without violating lepton number. But this non-zero vev is needed only for rank-6 models. In our discussion, we are free to choose either way, and we follow the former since rank-5 models are considered. In addition, inducing a negative mass for $\tilde{\nu}_L^c$ via the renormalization group may not always be  possible due to the necessity of large Yukawa couplings \cite{Campbell:1986xd,HR}. In the rest of our discussion we take $h_L$ as leptoquark ($W_3=0$), for reasons to be discussed shortly and call for the usual assignment to $\nu_L^c$. 

Now, the ALR and Inert models are defined as follows:
\begin{enumerate}
\item \underline{The ALR Model:}

The $SU(2)$ symmetry (the so-called $SU(2)_{R^{\prime}}$), $H^c\Leftrightarrow L,\,u_L^c\Leftrightarrow h_L^c,\,e_L^c\Leftrightarrow S_L^c$, imposed on the superpotential $W$, gives rise to the effective rank-5 version of ALR model and sets the following relations among $\lambda$'s: $\lambda_1=\lambda_7,\,\lambda_3=\lambda_4,\,\lambda_5=\lambda_6$.  Hence, by modifying the Yukawa couplings accordingly, the superpotential for the ALR model is written in a more compact form 
\begin{eqnarray}
W_{\rm ALR}&=&-\lambda_1 L_A^cF_AH+\frac{\lambda_2}{2}F_AF_A\nu_L^c +\lambda_3 QF_AX_A^c +\lambda_4 d_L^cQH\nonumber\\
&&+\lambda_5 h_L X_A^cL_A^c+\lambda_6 h_Ld_L^c\nu_L^c\nonumber\\
&=&\lambda_1\left(e_Le_L^cN_L-N_LN_L^cS_L^c+E_LE_L^cS_L^c-e_L^cE_L\nu_L\right)+\lambda_2\left(\nu_L\nu_L^cN_L^c-e_LE_L^c\nu_L^c \right)\nonumber\\
&&+\lambda_3\left(u_Lu_L^cN_L^c-d_Lu_L^cE_L^c+d_Lh_L^c\nu_L-u_Lh_L^ce_L \right)+\lambda_4\left(-d_Ld_L^cN_L+d_L^cu_LE_L\right)\nonumber\\
&&+\lambda_5\left(-h_Lh_L^cS_L^c+h_Lu_L^ce_L^c\right)+\lambda_6 h_Ld_L^c\nu_L^c,\label{SPALR}
\end{eqnarray}
where the following definitions are used in the first form of the above equation:
\begin{eqnarray}
F_A\equiv\left(H^c\;L\right)_L=\left(\begin{array}{cc}
E_L^c& \nu_L\\
N_L^c& e_L \\
\end{array}\right),\,L_A^c=\left(e^c\;S^c\right)_L,\,X_A^c=\left(h^c\;u^c\right)_L\,.
\end{eqnarray}
When the usual assignments of the Standard Model fields are taken, the baryon and the lepton numbers (with $R$-parity) of the exotics are $(B,L,R)_{H}=(0,0,-1),(B,L,R)_{S^c_L}=(0,0,-1)$. As discussed above, of the two possible assignments for $\nu_L^c$ in ALR model, the choice $(B,L,R)_{\nu^c_L}=(0,-1,1)$ is adopted. We also choose $h_L$ as a leptoquark $(B,L,R)_{h_L}=(1/3,1,-1)$. It should be noted that even though there are three possibilities for the assignments of the $h_L$ quantum numbers
 (leptoquark, di-quark or quark) at the $E_6$ level, $h_L$ is forced to be a leptoquark in both the  ALR and Inert models. This is simply because of the fact that the $SU(2)$ symmetries which convert $e_L^c\Leftrightarrow S_L^c$ and $\nu_L^c\Leftrightarrow S_L^c$ for the ALR and Inert models respectively would otherwise be broken. 

Thus, the superpartners of $N_L,N_L^c$ and $S_L^c$ $(\tilde{N}_L,\tilde{N}_L^c,\tilde{S}_L^c)$ are possible candidates which can play the role of the neutral Higgs fields. So the Higgs sector of the ALR model that we adopt is 
\begin{eqnarray}
H_1  = \left(\phi_1^+\;\tilde{S}_L^c\right),\;\; H_2 =  \left(\begin{array}{c}
\tilde{N}_L\\
\tilde{E}_L \\
\end{array}
\right),\;\; H_3 =\left(\begin{array}{cc}
\tilde{E}_L^c&\tilde{\nu}_L\\
\tilde{N}_L^c&\tilde{e}_L\\
\end{array}
\right),\;\;H_S =\phi_S^0.
\label{higgsALR}
\end{eqnarray}
Here the non-zero vev's are $\langle\tilde{S}_L^c\rangle = N_1,\,\langle \tilde{N}_L\rangle = v_1,\,\langle \tilde{N}_L^c\rangle = v_3,\,\langle \phi_S^0 \rangle = N_2$. In principle, one could have a non-zero vev for $\tilde{\nu}_L$, but this would violate lepton number (the neutrino, $\nu_L$, would get a large Majorana mass, through $\tilde{Z}_L$ exchange, of order $\langle \tilde{\nu}_L\rangle^2/M_{\tilde{Z}_L}$). We will assume that $\tilde{\nu}_L$ has no vev for the moment, but will mention the effects of the other possibility later. This other possibility could be acceptable if, for example, the $\tilde{\nu}_L$ comes from a different {\bf 27} than the Standard model fermions. One can choose $\phi_S^0$ to be a singlet or $\tilde{\nu}_L^c$. In the latter case, either $N_2$ needs to be zero or $(B,L,R)_{\nu^c_L}=(0,0,-1)$ should be adopted. Here, we keep our discussion general.

\item \underline{The Inert Model:}

In this case, the $SU(2)$ symmetry (the so-called $SU(2)_I$), $H\Leftrightarrow L,\,d_L^c\Leftrightarrow h_L^c,\,\nu_L^c\Leftrightarrow S_L^c$, imposed on the superpotential $W$ of Eq.~(\ref{SP}) leads to the effective rank-5 version of the Inert model. Thus, the following relations among the Yukawa couplings hold: $\lambda_2=\lambda_7,\,\lambda_4=\lambda_{11},\,\lambda_5=\lambda_8$. Similar to the ALR model case, the superpotential of the Inert model is expressed as 
\begin{eqnarray}
W_{\rm Inert}&=&\lambda_1^{\prime} L_I^cF_IH^c-\frac{\lambda_2^{\prime}}{2}F_IF_Ie_L^c +\lambda_3^{\prime}
h_LX^c_I L^c_I +\lambda_4^{\prime} h_Lu^c_L e_L^c+\lambda_5^{\prime} Q u^cH^c+\lambda_6^{\prime} QX^c_IF_I\nonumber\\
&=&\lambda_1^{\prime}\left(\nu_L\nu_L^cN_L^c+N_LN_L^cS_L^c+E_LE_L^cS_L^c+e_L\nu_L^cE_L^c\right)\nonumber\\
&&+\lambda_2^{\prime}\left(e_Le_L^cN_L+\nu_L\nu_L^cN_L^c+e_L^cE_L\nu_L \right)+\lambda_3^{\prime}\left(h_Lh_L^cS_L^c+h_Ld_L^c\nu_L^c\right)+\lambda_4^{\prime} h_Lu^c_L e_L^c\nonumber\\
&&+\lambda_5^{\prime}\left(u_Lu_L^cN_L^c+d_Lu_L^cE_L^c\right)\nonumber\\
&&+\lambda_6^{\prime}\left(d_Ld_L^cN_L+d_Lh_L^c\nu_L+u_Ld_L^cE_L+ u_Lh_L^ce_L\right)\,,\label{SPInert}
\end{eqnarray}
where the following definitions are used 
\begin{eqnarray}
F_I\equiv\left(H\;L\right)_L=\left(\begin{array}{cc}
N_L& \nu_L\\
E_L& e_L \\
\end{array}\right),\,L^c_I=\left(\nu^c\;S^c\right)_L
,\,X^c_I=\left(h^c\;d^c\right)_L.
\end{eqnarray}
The baryon and lepton number assignments for exotics are similar to the ALR model. $(B,L,R)_{H}=(0,0,-1)$ and $(B,L,R)_{S^c_L}=(0,0,-1)$ apply and $h_L$ is also considered as leptoquark as discussed above. Unlike the ALR case, $\nu_L^c$ is forced to have assignments $(B,L,R)_{\nu^c_L}=(0,-1,1)$. Thus, a vev for ${\tilde \nu}_L^c$ is not allowed unless lepton flavor violating interactions are included. From these considerations we choose the Higgs content of the model as follows
\begin{eqnarray}
H_D  = \left(\phi_S^0\;\tilde{S}_L^c\right),\;\;H_2  =  \left(\begin{array}{c}
\tilde{E}_L^c\\
\tilde{N}_L^c \\
\end{array}
\right),\;\; H_3 =\left(\begin{array}{cc}
\tilde{N}_L&\tilde{\nu}_L\\
\tilde{E}_L&\tilde{e}_L\\
\end{array}
\right),\;\;H_S=\phi_1^+,
\label{higgsInert}
\end{eqnarray}
with the following vev's, $\langle \tilde{S}_L^c\rangle = N_1,\,\langle \tilde{N}_L\rangle = v_1,\,\langle \tilde{N}_L^c\rangle = v_3,\,\langle \phi_S^0 \rangle = N_2$. Here the $SU(2)_I$ doublet $H_D$ is electrically neutral while $\phi_S^0$ is possibly taken as $\tilde{\nu}_L^c$ with zero vev. As before, we assume that $\langle \tilde{\nu}_L\rangle =0$, but will consider the alternative possibility later.
\end{enumerate}

\section{Neutrinos in the ALR and Inert Models}\label{sec:ALRI}

In this section we analyze the neutral fermion sectors of both the ALR and Inert models\footnote{Since this paper is concentrating 
on neutrinos, we will not discuss mixing between light and heavy fields in the charged 
lepton or quark sectors.  Such mixing can have a wide 
range of interesting phenomenological effects, see Ref. \cite{Frampton:1999xi} 
for a detailed discussion and a 
list of references.} by using the superpotentials given in Eqs. (\ref{SPALR}) and (\ref{SPInert}). The superpotentials only describe $({\bf{27}})^3$ type interactions. Without considering any more particles or new interactions, there exists a $5\times 5$ ``neutrino" mass matrix for each generation. From $R$-parity considerations this $5\times 5$ matrix splits into $2\times 2$ and $3\times 3$ submatrices. From Eqs. (\ref{SPALR}) and (\ref{SPInert}), the $R=+1$ neutral fermion sector spanned by $(\nu_L,\,\nu_L^c)$ becomes
\begin{eqnarray}
{\cal M}^{R=+1}=\left(\begin{array}{cc}
0            & m_{\nu\nu^c}\\
m_{\nu\nu^c} & 0 
\end{array}\right),
\end{eqnarray}
where $m_{\nu\nu^c}=\lambda_2 \langle \tilde{N}_L^c\rangle=\lambda_2 v_3$ in the ALR and 
$m_{\nu\nu^c}^{\prime}=\lambda_1^{\prime} v_3$ in the Inert model. Clearly, the ordinary neutrinos have a Dirac mass 
$m_{\nu\nu^c}$ which is of the order of the up quark mass in both models and the physical state is formed
by the maximal mixing of $\nu_L$ and $\nu_L^c$. Either an unnatural fine tuning for the Yukawa couplings is needed, or we must introduce a large Majorana mass for $\nu_L^c$ which  renders a 
small Majorana  mass for $\nu_L$ through the canonical seesaw mechanism \cite{Mohapatra:1979ia}. Another 
possibility is to generate a small Dirac one-loop mass by eliminating the tree level mass term. The possibilities will be discussed shortly.

The $R=-1$ sector is composed of 3 neutral leptons, $N_L,N_L^c,S_L^c$, and 3 neutral gauge fermions 
corresponding to two the $SU(2)$'s and one $U(1)$ group. For simplicity, we assume that the gauge fermions 
get large Majorana mass terms from soft-supersymmetry breaking and decouple. The $3\times 3$ Majorana mass matrices in the  $(N_L,N_L^c,S_L^c)$ basis become
\begin{eqnarray}
{\cal M}_{\rm ALR}^{R=-1}=\left(\begin{array}{ccc}
0              & - m_{EE^c}  &  -\lambda_1 v_3  \\
    -m_{EE^c}  &  0          &  -m_{ee^c}      \\
-\lambda_1 v_3  & -m_{ee^c}   &  0       
\end{array}\right),\;
{\cal M}_{\rm I}^{R=-1}=\left(\begin{array}{ccc}
0                     & m_{EE^c}^{\prime}  &  m_{\nu\nu^c}^{\prime}  \\
 m_{EE^c}^{\prime}    &  0                 &  m_{ee^c}^{\prime}          \\
m_{\nu\nu^c}^{\prime} & m_{ee^c}^{\prime}  &  0       
\end{array}\right),
\end{eqnarray}
where $m_{EE^c}=\lambda_1 N_1,\,m_{ee^c}=\lambda_1 v_1,\,m_{EE^c}^{\prime}=\lambda_1^{\prime} N_2,\,m_{ee^c}^{\prime}=\lambda_1^{\prime} v_1$. Here ${\cal M}_{\rm ALR} ({\cal M}_{\rm I})$ is the $R=-1$ mass matrix in  the ALR (Inert) model. 
Diagonalization of the above matrix for the ALR case leads to the states and masses
\begin{eqnarray}
|\nu_{1,2}\rangle_{\rm ALR} &\simeq & \frac{1}{\sqrt{2}}\left(|N_L\rangle \pm |N_L^c\rangle\right),\;\;M_{1,2}^H\simeq\pm m_{EE^c},\nonumber\\
|\nu_3\rangle_{\rm ALR} &\simeq & |S_L^c\rangle,\hspace*{2.48cm}M_3^L\simeq 2\lambda_1 v_3 m_{ee^c}/m_{EE^c},
\end{eqnarray}
under the assumption $m_{ee^c}\sim\lambda_1v_3\ll M_{EE^c}$. Here the superscripts $H$ and $L$ stand for the heavy and light states, and $\lambda_1$ can further be express as $m_{EE^c}/N_1$. 
Similar results apply to the Inert model, where the states are the same with masses $M_{1,2}^{\prime H}=\pm m_{EE^c}^{\prime}$ and $M_3^{\prime L}=-2 m_{\nu\nu^c}^{\prime}m_{ee^c}^{\prime}/m_{EE^c}^{\prime}$, respectively. Clearly, there are two heavy ($|\nu_{1,2}\rangle_{\rm ALR(I)}$) states and one light ($|\nu_3\rangle_{\rm ALR(I)}$) state. The light (mainly sterile) state  does not however mix with active neutrinos 
unless $R$-parity is broken. In the non-SUSY framework \cite{Frank:2004vg}, without introducing further symmetries or interactions the lightest state is formed by $\nu_L^c$ and $S_L^c$ . 
In the SUSY scenario, it is still possible to mix $R=+1$ and $R=-1$ sectors by including soft-symmetry breaking terms \cite{ma92}. This will be discussed in the next section. 

At this stage, one can take $(\tilde{\nu}_L\;\tilde{e}_L)$ with non-zero vev for $\tilde{\nu}_L$ without changing the above results. However, for the case $\phi_S^0 = \tilde{\nu}_L^c$ with zero vev,  the results are modified for the Inert model but remain unchanged for ALR. Then, the $R=-1$ sector of the Inert model has the following states and masses
\begin{eqnarray}
|\nu_{1,2}^{\prime}\rangle_{\rm I} \,&\stackrel{\langle \tilde{\nu}_L^c\rangle=0}{\longrightarrow}&\,\frac{1}{\sqrt{2}}\left(\frac{m_{\nu\nu^c}^{\prime}}{M_1^{\prime H}} |N_L\rangle +  \frac{m_{ee^c}^{\prime}}{M_1^{\prime H}}|N_L^c\rangle \pm |S_L^c\rangle \right),\;M_{1,2}^{\prime H}\,\stackrel{\langle \tilde{\nu}_L^c\rangle=0}{\longrightarrow}\,\pm\sqrt{m_{\nu\nu^c}^{\prime 2}+m_{ee^c}^{\prime 2}},\nonumber\\
|\nu_3^{\prime}\rangle_{\rm I} \,&\stackrel{\langle \tilde{\nu}_L^c\rangle=0}{\longrightarrow}&\,\frac{1}{M_1^{\prime H}}\left(-m_{ee^c}^{\prime}|N_L\rangle + m_{\nu\nu^c}^{\prime}|N_L^c \rangle \right),\;\hspace*{1.3cm}M_3^{\prime L}\,\stackrel{\langle \tilde{\nu}_L^c\rangle=0}{\longrightarrow}\,0,\label{nucInert}
\end{eqnarray}

For the case  $\phi_S^0 = \tilde{\nu}_L^c$ with non-zero vev (that is, when $(B,L,R)_{\nu^c_L}=(0,0,-1)$ is adopted), the $R=-1$ sector of the ALR model would be a $4\times 4$ matrix. This possibility is solely available for the ALR model, since $R$-parity conservation requires $\lambda_1^{\prime}$ in Eq.~(\ref{SPInert}) to vanish. In ALR, $\lambda_2$ should be eliminated by imposing some discrete symmetries in order not to break $R$-parity conservation. However, this also decouples $\nu^c_L$ from the $4\times 4$ matrix and makes it massless. So, no change occurs in the $3\times 3$ submatrix and both $\nu_L$ and $\nu_L^c$ become massless. Note that in this framework $\nu_L^c$ is no longer a Dirac conjugate pair state  of the active $\nu_L$ neutrino but it is a sterile neutrino with zero lepton number. In the next section, we discuss possible mechanisms to generate small Majorana masses for active neutrinos and possible mixing between opposite $R$-parity sectors.

\section{Giving Mass Through $R$-parity Breaking}\label{sec:Rviol}

The fact that the $R$-parity might be broken by soft terms \cite{ma92} has been discussed by Ma in the context 
of ALR model \cite{Ma:2000ww}. The idea is as follows. A soft term which describes a mixing between $\nu_L$ and $N_L^c$ can be realized by, for example, giving a vev to ${\tilde \nu}_L^c$ in the $F_AF_A\nu_L^c$ term of Eq.~(\ref{SPALR}). It can be defined as $\mu_A(\nu_LN_L^c-e_LE_L^c)$. The presence of such mixing then induces a mixing between $\nu_L$ and the lightest state $\nu_3$ through a small $N_L^c$ component of $\nu_3$. Then, the active neutrino mass matrix is enlarged from $2\times 2$ to $3\times 3$ and is given by, in the basis $\left(\nu_L,\nu_L^c,S_L^c\right)$ \footnote{The third entry will be represented by $S_L^c$ since the lightest state is mainly described by $S_L^c$.}
\begin{eqnarray}
{\cal M}_{\rm ALR}^{R=+1}=\left(\begin{array}{ccc}
0             &   m_{\nu\nu^c}  &  m_S \\
m_{\nu\nu^c}  &       0         &    0 \\
m_S           &       0         &   M_3^L 
\end{array}\right).\label{mixedmatrix}
\end{eqnarray}
where $m_S\equiv m_{\nu\nu^c}\mu_A/{m_{EE^c}}$ and $M_3^L\simeq 2\lambda_1 v_3 m_{ee^c}/m_{EE^c}$.
The corresponding matrix for Inert model is ${\cal M}_{\rm I}^{R=+1}={\cal M}_{\rm ALR}^{R=+1}\left(m_{\nu\nu^c}(M_3^L)\to m_{\nu\nu^c}^{\prime}(M_3^{\prime L}), m_S\to m_S^{\prime}\equiv m_{\nu\nu^c}^{\prime}\mu_I/m_{EE^c}^{\prime}\right)$. Here $M_3^{\prime L}$ is defined as $M_3^{\prime L}=-2 m_{\nu\nu^c}^{\prime}m_{ee^c}^{\prime}/m_{EE^c}^{\prime}$
We envisage two limiting cases,  one for $\mu_A$ very small compared with $m_{EE^c}$; $m_S\ll|M_3^L|\ll m_{\nu\nu^c}$ (case~(i)) and one for $\mu_A$ large compared with $m_{EE^c}$; $|M_3^L|\ll m_S\ll m_{\nu\nu^c}$ (case~(ii)). A third case is possible when $\mu_A$ is comparable with $m_{EE^c}$; $|M_3^L|\ll m_S\sim m_{\nu\nu^c}$ (case~(iii)). In the case in which all three, $|M_3^L|, m_S$ and $m_{\nu\nu^c}$ are comparable with each other, it is not possible to draw any valuable conclusion from analytic calculations. In order get sizable mixing between active and sterile neutrinos while they are lying in the correct mass range, it is required to have 
comparable but small Dirac and Majorana masses \cite{Langacker:1998ut}. The main results of the above cases and each of the corresponding spectra are summarized in Table~\ref{Rviolation} in Appendix. 

Summarizing the results from Appendix, one finds that the spectrum has two heavy and one light states. The common feature of all cases is that the lightest state is always purely sterile, mainly composed of either $S_L^c$ or $S_L^c$ and $\nu_L^c$. It has a seesaw type mass as given in Appendix. The heavy states have maximal mixing between $\nu_L$ and $\nu_L^c$ with a small component of $S_L^c$. For case~(i) ($m_S\ll|M_3^L|\ll m_{\nu\nu^c}$), the heavy states consist of only $\nu_L$ and $\nu_L^c$. They are too heavy to be considered the physical neutrino state as $m_{\nu\nu^c}$ is at the scale of the up quark mass.

Thus so far no satisfactory pattern for neutrino masses and mixings has been established. We are still required to go beyond the minimal picture, as we will discuss in the next section. 

If the $\tilde{\nu}_L^c$ comes from a different ${\bf 27}$, and thus can get a non zero vev, then these results are unaffected. Choosing $\phi_S^0 = \tilde{\nu}_L^c$ with zero vev, however, makes $R$-parity violation disappear and the $S_L^c$ again decouples. 

\section{Beyond the Minimal Content}\label{sec:mixing}

As we have seen in the previous sections, the absence of the terms in $W_3$, which guarantees proton stability, could be a consequence of a discrete symmetry. Note that conventional $R$-parity is not sufficient to explain the elimination of some Yukawa couplings. For this purpose, an odd $Z_2$ charge to all Standard Model quarks and $h_L$ and 
an even $Z_2$ charge to the rest of the fields can be assigned. Clearly, invariance under this $Z_2$ symmetry would require $\lambda_9$ and $\lambda_{10}$ to be zero. Elimination of other Yukawa couplings can be achieved by imposing further symmetries. Depending on whether the neutrinos are Dirac or Majorana particles, we can proceed in two ways. 

If one assumes that the neutrino is a Dirac particle, then the Dirac neutrino mass predicted directly from the superpotential should be much smaller for both models. A solution to effectively fine tune the coupling has been proposed by Branco and Geng \cite{Branco:1987yg, Branco:1987ab}. They make the model invariant under a $Z_3$ symmetry in addition to the $Z_2$ symmetry considered above (this is what we have called the Discrete Symmetry (DS) method in our earlier paper \cite{Frank:2004vg}). Here the $Z_3$ symmetry distinguishes between generations. The symmetry eliminates the tree-level Dirac mass term from the superpotential and induces a smaller one-loop mass. In Ref.~\cite{Branco:1987yg}, the discussion has been carried out at $E_6$ level without reference to any of its subgroups. Assuming the invariance of $E_6$ 
itself under $Z_3$ symmetry, the breaking of $E_6$ to the $SU(2)_{R^{\prime}}$ or $SU(2)_I$ symmetries lead to breaking of the $Z_3$ symmetries. Since the ALR and Inert models are treated as different subgroups of $E_6$, one can introduce $Z_3$ invariance after the $E_6$ gauge symmetry is broken.

If neutrinos are considered to be Majorana particles, then generation of small Majorana masses for left-handed (active) neutrinos could be achieved by including Higher Dimensional Operators (HDO) \cite{Nandi:1985uh}. One can show that 
the next available interactions in the Standard Model  are dimension-5, which can be sizable if one introduces an intermediate scale around $10^{11}$ GeV and Higgs fields from a ${\bf \overline{27}}$ representation of $E_6$. Through the canonical seesaw mechanism, in the   $R=+1$ sectors of the models, the small Majorana mass of the left-handed active neutrino is generated by having a large Majorana mass for $\nu_L^c$. In the $R=-1$ sector, $S_L^c$ will also get a large Majorana mass which modifies the results discussed in Section \ref{sec:ALRI}. If one further includes the soft-terms which break $R$-parity, a large coupling could occur between $\nu_L^c$ and $S_L^c$. So, this framework can give us a picture involving sterile neutrinos, which is promising. 

As an alternative to the above methods, one can extend the minimal content of $E_6$ together with its  Higgs sector by further considering $E_6$-neutral fermion and Higgs fields from the split multiplet ${\bf 27} + {\bf \overline{27}}$. This Additional Neutral Fermion (ANF) method was first proposed by Mohapatra and Valle \cite{Mohapatra:1986bd,Mohapatra:1986ks,Mohapatra:1986aw}. This way, it is possible to produce either the Dirac or Majorana neutrinos with small mass.

Among these methods, the DS is the simplest and the most attractive one as it does not require the existence of an intermediate scale or inclusion of new particles (and interactions). The ANF method is the most complex, as it requires  not only presence of some  discrete symmetries but also the presence of new interactions. As indicated earlier, the occurrence of sterile neutrino components in the physical states can only be possible through soft-breaking terms. We now analyze 
these methods in the following subsections.

\subsection{The Discrete Symmetry Method}\label{subsec:DS}

As discussed above, a $Z_2$ symmetry which assigns odd charges to $Q,u^c_L,d_L^c,h_L,h_L^c$ fields and even charges to the rest may be required to explain the absence of $W_3$-terms. The DS method, as we will show, also imposes a $Z_3$ symmetry which eliminates the tree-level $m_{\nu\nu^c}$ and makes it appear at one-loop. It is thus much smaller. Unlike the Inert case, in the ALR model $M_3^L$ does not depend on $m_{\nu\nu^c}$. So, one-loop Dirac mass generation for neutrinos doesn't affect $M_3^L$ and make it comparable or even bigger than the $m_{\nu\nu^c}$ generated at one-loop (say $m_{\nu\nu^c}^{\rm 1-loop}$) in ALR. The cases for ALR should thus  be reconsidered under the circumstance $m_S\ll m_{\nu\nu^c}^{\rm 1-loop}\sim M_3^L$. As a result case~(ii) is irrelevant and in case~(iii) the hierarchy among $|M_3^L|, m_{\nu\nu^c}$, and $m_S$ disappears. So, three of the parameters become comparable with each other and no conclusion can be extracted in this case. 

\begin{itemize}
\item \underline{One-Loop Masses in ALR:}

In addition to the $Z_2$ symmetry, a $Z_3$ symmetry is needed to set $\lambda_2$ to zero. It should of course leave the Yukawa couplings $\lambda_3$ and $\lambda_6$ in Eq.~(\ref{SPALR}) unaffected to generate one-loop neutrino mass, and $\lambda_1,\lambda_4$ and $\lambda_5$ to generate masses for Standard Model quarks and charged leptons and exotics. This should be the case for at least some components of these couplings in flavor space. We know that the $Z_3$ symmetry should distinguish between generations \cite{Branco:1987yg}. One of such allowed symmetry charge assignments could be as follows:
\begin{eqnarray}   
Z_3:&&\left[Q,d^c_L,h_L,h_L^c,L,\nu_L^c\right]^{(i)}\to \eta \left[Q,d^c_L,h_L,h_L^c,L,\nu_L^c\right]^{(i)}\,,\nonumber\\
&& u_L^{c(i)} \to \eta^{-1} u_L^{c(i)}\,,\nonumber\\
&&H^{ c (1)}\to \eta^{-1}H^{ c (1)}\,,\;\; H^{ c (2)}\to H^{ c (2)}\,,\;\;\; H^{ c (3)}\to H^{ c (3)}\,,\nonumber\\
&&H^{ (1)}\to \eta^{-1}H^{ (1)}\,,\;\;\;\;\;\,H^{ (2)}\to \eta H^{ (2)}\,,\;\;\;\, H^{ (3)}\to H^{ (3)}\,,\nonumber\\
&&S_L^{c(1)}\to \eta^{-1}S_L^{c(1)}\,,\;\;\;\; S_L^{c(2)}\to S_L^{c(2)}\,,\;\;\;\;\; S_L^{c(3)}\to \eta S_L^{c(3)},
\end{eqnarray}
where $\eta^3=1$ and the numbers inside the parentheses represent generations. The masses for quarks, charged leptons and exotics are given as
\begin{eqnarray}
m_u&=&\lambda_3^{ij3}\langle\tilde{N}^{c (3)}\rangle,\;\;m_d=\lambda_4^{ij2}\langle\tilde{N}^{(3)}\rangle,\;\;\;\;m_e=\lambda_1^{ij1}\langle\tilde{N}^{(1)}\rangle,\nonumber\\
m_h&=&\lambda_5^{ij3}\langle\tilde{S}^{c (3)}\rangle,\;\;m_E=\lambda_1^{ij1}\langle\tilde{S}^{c (1)}\rangle,\;\;\;m_N=\lambda_1^{ij1}\langle\tilde{S}^{c (1)}\rangle.
\end{eqnarray}
\begin{figure}
\vspace*{-4cm}
\begin{center}
\includegraphics[height=8.9in,width=6.9in,angle=0]{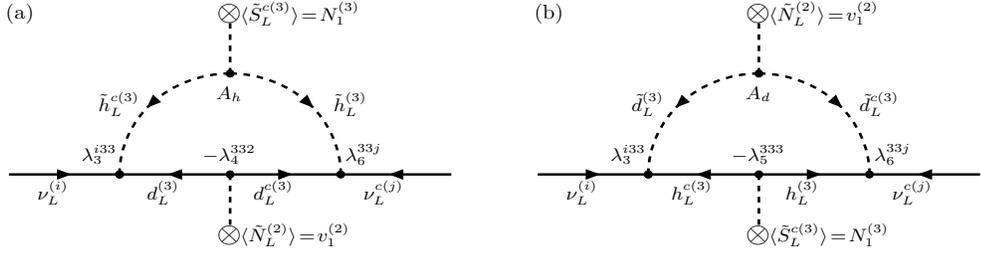}
\vspace*{-15.5cm}
\caption{\label{fig:SUSYdiracmass}
The one-loop Dirac masses for $\nu_L^{(i)}\nu_L^{c(j)}$.}
\end{center}
\end{figure}
There are two one-loop diagrams as shown in Fig.~\ref{fig:SUSYdiracmass} which contribute to the Dirac mass generation for $\nu_L$. A trilinear scalar vertex is involved in the one-loop diagrams. We take $A_h$ and $A_d$ as the trilinear soft supersymmetry-breaking coefficients for $\tilde{h}_L\tilde{h}_L^c\tilde{S}_L^c$ and $\tilde{d}_L\tilde{d}_L^c\tilde{N}_L$, respectively. Only $\lambda_3,\lambda_4,\lambda_5$ and $\lambda_6$ are involved in the one-loop diagrams. If we take the mass of $h_L$ as the typical SUSY breaking scale $m_{1/2}$, then the one-loop neutrino mass is obtained by adding two diagrams in Fig.~\ref{fig:SUSYdiracmass} 
\begin{eqnarray}
m_{\nu\nu^c}^{\rm 1-loop}&=&m_{\nu\nu^c}^{\rm 1-loop (a)}+m_{\nu\nu^c}^{\rm 1-loop (b)}\nonumber\\
&\simeq&\frac{A_h \lambda_3^{i33}\lambda_6^{33j}m_b}{32\pi^2}
\end{eqnarray}
where we have assumed the soft supersymmetry-breaking squark masses participating in the one-loop diagrams are given as \cite{Campbell:1986xd} $m_{\tilde{d}}\sim m_{\tilde{h}}\simeq 3 m_{1/2}$ and $A_h=A_d$. In order to obtain neutrino masses less than $0.1$ eV, a bound $\lambda_3^{i33}\lambda_6^{33j}\leq 7\times 10^{-9}$ must be imposed for all $i,j$ when $A_h$ is taken of order one. This is not substantially smaller than typical Yukawa couplings.

As discussed above, having a Dirac mass for neutrinos less than an eV requires reconsideration of the case~(i) of section \ref{sec:Rviol} under the new hierarchy $m_S\ll m_{\nu\nu^c}^{\rm 1-loop}\sim M_3^L$ and makes the other cases irrelevant or inconclusive. It is not possible to give some useful analytic expressions for the masses and states unless a specific relation between $M_3^L$ and $m_{\nu\nu^c}^{\rm 1-loop}$ is set. For illustrative purposes, if $M_3^L=3m_{\nu\nu^c}^{\rm 1-loop}$ is chosen, the physical states under the assumption $m_S\ll  M_3^L=3m_{\nu\nu^c}^{\rm 1-loop}$ become
\begin{eqnarray}
|\nu_{1}\rangle&\simeq& \frac{1}{\sqrt{2}}\left(|\nu_L\rangle + |\nu_L^c\rangle -\frac{\xi}{2} |S_L^c\rangle \right),\;M_1\simeq  m_{\nu\nu^c}^{\rm 1-loop},\nonumber\\
|\nu_{2}\rangle&\simeq& \frac{1}{\sqrt{2}}\left(|\nu_L\rangle - |\nu_L^c\rangle -\frac{\xi}{4} |S_L^c\rangle \right),\;M_2\simeq - m_{\nu\nu^c}^{\rm 1-loop},\nonumber\\
|\nu_3\rangle &\simeq& \frac{3\xi}{8}|\nu_L\rangle +\frac{\xi}{8}|\nu_L^c\rangle +|S_L^c\rangle,\hspace*{1.1cm}M_3\simeq 3m_{\nu\nu^c}^{\rm 1-loop},
\label{mixcase3DS}
\end{eqnarray}
where now $\xi\equiv m_S/m_{\nu\nu^c}^{\rm 1-loop}$ is implied. We still have two states $\nu_{1,2}$ showing a bi-maximal mixing between the active $\nu_L$ neutrino and $\nu_L^c$ where as the sterile state  $S_L^c$ appears as separate. Since the masses lie in the acceptable range, it would be possible to obtain {\bf{3}+{\bf 2}} structural models.

\item\underline{One-Loop Masses in Inert:}

There are some differences between the two models in terms of the required $Z_3$ charge assignments. Firstly, 
since both $m_S^{\prime}$ and $M_3^{\prime L}$ depend on $m_{\nu\nu^c}^{\prime}$ linearly in the Inert case, the DS method 
doesn't change the hierarchy among them. So, there are only the three cases as discussed in section \ref{sec:Rviol} after the discrete symmetry is imposed. Secondly, an even $Z_3$ charge can be assigned to $u_L^c$ for three generations since the $\lambda_6$ term of Eq.~(\ref{SP}) is invariant under $SU(2)_I$ and it can be eliminated by $Z_3$ invariance without leading to any problems. Lastly and most importantly, unlike the ALR case, the $\lambda_{11}$ term of Eq.~(\ref{SP}) is not invariant under $SU(2)_I$ symmetry and is combined with the $\lambda_4$ term (we relabeled both as $\lambda_1^{\prime}$ in Eq.~(\ref{SPInert})). Thus, it is not possible to eliminate the tree-level Dirac mass term ($\lambda_{11}$) for active neutrinos unless some further assumptions are made, since eliminating the $\lambda_{11}$ term would also eliminate the mass terms for $N_L$ and $E_L$. 

The assumption needed could be to take the vev's of $\tilde{H}^{c}$ zero for the first two generations and giving mass to the up-quarks from $\tilde{H}^{c (3)}$ whose vev is assumed non-zero. Then the following charges could be assigned to the fields
\begin{eqnarray}   
Z_3:&&\left[Q,d^c_L,h_L,h_L^c,L,\nu_L^c\right]^{(i)}\to \eta \left[Q,d^c_L,h_L,h_L^c,L,\nu_L^c\right]^{(i)}\,,\nonumber\\
&&H^{ c (1)}\to \eta H^{ c (1)}\,,\;\;\; H^{ c (2)}\to \eta H^{ c (2)}\,,\;\; H^{ c (3)}\to \eta^{-1}H^{ c (3)}\,,\nonumber\\
&&H^{ (1)}\;\to \eta H^{ (1)}\,,\;\;\;\;\;\, H^{ (2)}\to \eta H^{ (2)}\,,\;\;\;\;\; H^{ (3)}\to \eta^{-1} H^{ (3)}\,,\nonumber\\
&&S_L^{c(1)}\to \eta^{-1}S_L^{c(1)}\,,\;\; S_L^{c(2)}\to S_L^{c(2)}\,,\;\;\;\;\;\;S_L^{c(3)}\to \eta S_L^{c(3)},
\end{eqnarray}
where 
\begin{eqnarray}
m_u&=&\lambda_5^{\prime ij3}\langle\tilde{N}^{c (3)}\rangle,\,\;\;\;\;\;\;\;\;\;m_d=\lambda_6^{\prime ij\alpha}\langle\tilde{N}^{(\alpha)}\rangle,\,\;\;\;\;\;\;\;\;m_e=\lambda_2^{\prime ij3}\langle\tilde{N}^{(3)}\rangle,\nonumber\\
m_h&=&\lambda_3^{\prime ij3}\langle\tilde{S}^{c (3)}\rangle,\;\;\;\;m_{E^{\alpha},N^{\alpha}}=\lambda_1^{\prime ij3}\langle\tilde{S}^{c (3)}\rangle,\,\,\;m_{E^3,N^3}=\lambda_1^{\prime ij1}\langle\tilde{S}^{c (1)}\rangle.
\end{eqnarray}
Here $\alpha$ runs over the first and the second generations. Then, one-loop diagrams giving non-zero Dirac neutrino masses are shown in Fig. \ref{fig:InertSUSYdiracmass}.
\begin{figure}
\vspace*{-4cm}
\begin{center}
\includegraphics[height=8.9in,width=6.9in,angle=0]{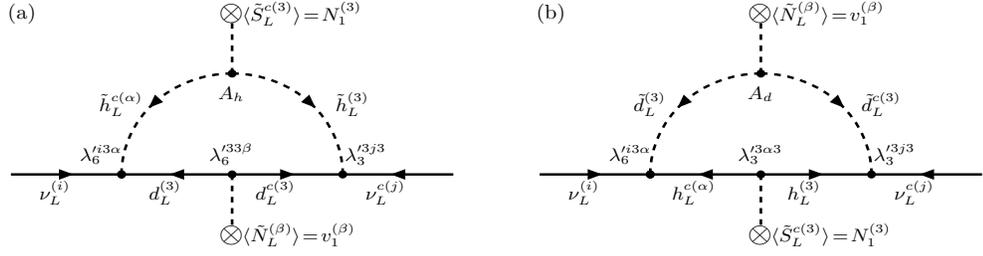}
\vspace*{-15.5cm}
\caption{\label{fig:InertSUSYdiracmass}The one-loop Dirac masses for $\nu_L^{(i)}\nu_L^{c(j)}$. The indices $\alpha$ and $\beta$ run over first two generations.}
\end{center}
\end{figure}
Under the same assumptions as in ALR case for the calculation of the one-loop integrals, we get 
\begin{eqnarray}
m_{\nu\nu^c}^{\prime \rm 1-loop}&=&m_{\nu\nu^c}^{\prime\rm 1-loop (a)}+m_{\nu\nu^c}^{\prime\rm 1-loop (b)}\nonumber\\
&\simeq&\frac{A_h \lambda_3^{\prime 3j3}\lambda_6^{\prime i3\alpha}m_b}{32\pi^2}
\end{eqnarray}
where $\alpha=1,2$. To get a Dirac mass $m_{\nu\nu^c}^{\rm \prime 1-loop}<0.1$ eV we need to impose the bound $\lambda_3^{\prime 3j3}\lambda_6^{\prime i3\alpha}<7\times10^{-9}$ for $\alpha=1,2$. Here, unlike the ALR case, it is possible to set separate bounds on $\lambda_3^{\prime 3\alpha 3}$ and $\lambda_6^{\prime 33\beta}$ using fact that they give masses to the $h_L$ and bottom quark, respectively. These bounds, however, become weaker. Having Dirac mass in the eV range makes the mostly sterile state $|\nu_3\rangle$ in all three cases too light to be detected. The other two states in each case have the large mixing problem. As in the non-SUSY case, the DS method is only able to explain the smallness of the Dirac neutrino mass and not the mixing.
\end{itemize}

\subsection{The Higher Dimensional Operators Method}\label{subsec:HDO}

This method adds higher dimensional interactions to the Lagrangian, which can substantially modify some of the fermion mixings. Due to the compactification scale suppression factor ($\sim 10^{19}$ GeV), it is sufficient to consider only dimension-5 interactions. The method also requires the existence of intermediate scales set by  some $SU(2)_L$ singlet Higgs fields from ${\overline{\bf{27}}}$ representation of $E_6$. So, there will be $\tilde{S}_L^c$-like and $\phi_S^0$-like scalars ($H_1$ and $H_S$) for the ALR model and an $H_D$-like Higgs doublet for the Inert model.\footnote{We will assume that all Higgs fields of ${\bf 27}$ whose vev's are at the electroweak scale have corresponding $\overline{\bf 27}$ Higgs fields with vev's at the same scale. All Higgs fields from the $\overline{\bf 27}$ representation will have opposite quantum numbers with respect to the fields in the ${\bf 27}$.} The vev's of these fields are written as $\langle\overline{\tilde{S}_L^c}\rangle=\Lambda_1$ and $\langle\overline{\phi_S}\rangle=\Lambda_2$. Here $\phi_S$ could be replaced with $\tilde{\nu}_L^c$. The dimension-5 interactions can be written as
\begin{eqnarray}
{\cal L^{\rm (5)}_Y}=\frac{f}{M_c}\psi^T({\bf{27}})\,\epsilon\,H(\overline{{\bf{27}}})\,C\, 
H^T(\overline{{\bf{27}}})\,\epsilon\,\psi({\bf{27}})\,,
\label{dim5Lag}
\end{eqnarray}
where $M_c$ is the compactification scale ($\sim 10^{19}$ GeV) and the Higgs field $H({\bf{27}})$ stands for both $\overline{\tilde{S}_L^c}$ and $\overline{\phi_S}$. Here $C$ is  the charge conjugation matrix defined as $C=\left( \begin{array}{cc}-\epsilon&0\\0 & \epsilon\\
\end{array}\right)$ and we adopt the chiral representation and $\epsilon\equiv i\sigma_2$, where $\sigma_2$ is the Pauli matrix.

The new interactions do not sizably modify any interaction terms in the fermion mass matrices with the exception of the $\nu_L^c-S_L^c$ submatrix. The matrix in the $(\nu_L,\nu_L^c,S_L^c)$ basis then becomes 
\begin{eqnarray}
{\cal M}^{5}=\left(\begin{array}{ccc}
0            &  m_{\nu\nu^c}  & m_S \\
m_{\nu\nu^c} &   K_1          & K_{12}\\
m_S          &   K_{12}       & K_2
\end{array}
\right)\,,
\label{HDOmassmatrix}
\end{eqnarray}
where $K_{i}\equiv f\frac{\Lambda_i^2}{M_c}$, $i=1,2$ and $K_{12}\equiv 2\sqrt{K_1 K_2}$. We keep the discussion in this section general and applicable to both the ALR and Inert models unless stated. Furthermore we note that the $M_3^L$ term in (3,3) entry of the above matrix is negligible with respect to $K_2$. We now consider three cases: (a) the case in which there is only one intermediate scale of the order of $10^{12}$ GeV (i.e. $\Lambda_1=\Lambda_2$), (b) the case in which $\Lambda_1$ is much smaller than $\Lambda_2$ and is of the order of $1$ TeV and (c) the case in which $\Lambda_2$ is much smaller than $\Lambda_1$ and is of the order of $1$ TeV.

Case~(a) with $\Lambda_1=\Lambda_2\sim 10^{12}$ GeV makes 
$K_1$ and $K_2$ (and thus $K_{12}$) much bigger than the other entries of the matrix and of the order of $10^{5}$ GeV. Here we are assuming the coupling constant $f$ is of the order of unity. With big Majorana masses for $\nu_L^c$ and $S_L^c$, these states decouple. One light and two heavy physical states are expected. Under the assumption $m_S\ll m_{\nu\nu^c}\ll K_1=K_2$ the states are
\begin{eqnarray}
|\nu_{1,2}\rangle&\simeq& \frac{1}{\sqrt{2}}\left(|\nu_L^c\rangle \pm |S_L^c\rangle \right),\;\;\;\;\;\;\;\;\;\overline{M_{1,2}^H}(K_1=K_2)\simeq  3K_1,-K_1,\nonumber\\
|\nu_{3}\rangle&\simeq& |\nu_L\rangle +\frac{\zeta}{3} |\nu_L^c\rangle -\frac{2\zeta}{3} |S_L^c\rangle,\;\;\;\overline{M_3^L}(K_1=K_2)\simeq \frac{1}{3}\zeta^2 K_1,
\label{HDOstates1}
\end{eqnarray}
where $\zeta\equiv \frac{m_{\nu\nu^c}}{K_1}$ and is of order $10^{-8}$, and the orthonormality of the states is guaranteed up to $O(\zeta)$ since $\zeta^2$ and higher terms are not included. The $\nu_L^c$ and $S_L^c$ components of $\nu_3$ are shown because their mixing with $\nu_L$ is order of $\zeta \sim 10^{-8}$, which is too small to be relevant. We get a seesaw-like small Majorana mass $\overline{M_3^L}(K_1=K_2)\simeq \frac{1}{3}\zeta^2 K_1$ for $\nu_L$, which is around $3\times 10^{-3}$ eV. Thus, consideration of case~(a) with $\Lambda_1=\Lambda_2\sim 10^{12}$ GeV gives rise to a single state having acceptable eV range mass with negligible active-sterile mixing. Furthermore, $S_L^c$ and $\nu_L^c$ appear as two distinct flavor states in $\nu_3$. 

The above results show that, in order to have two light states with significant active-sterile mixing, there should be a substantial hierarchy between $K_1$ and $K_2$ in order that only one of the states $|\nu_{1,2}\rangle$ decouples, which leaves two light states. We now consider case~(b) in which $\Lambda_1$ is much smaller than $\Lambda_2$. This also leads to a large $\nu_L$ Majorana mass since $m_{\nu\nu^c}$ is fixed by the model to be of order $1$ MeV. Thus this case is not realistic. However, one can consider the opposite case, case~(c), where $\Lambda_2$ is  much smaller than $\Lambda_1$ so that $K_2$ will be much smaller than both $K_1$ and $m_{\nu\nu^c}$.\footnote{$K_2$ may be comparable with $m_S$ but still big enough to neglect the $M_3^L$ term in (3,3) entry of the mass matrix in Eq.~(\ref{HDOmassmatrix}).} In this case the approximate states and masses become
\begin{eqnarray}
|\nu_{1}\rangle&\simeq& |\nu_L^c\rangle,\hspace{2cm}\overline{M_1^H}(K_2\ll K_1)\simeq  K_1,\nonumber\\
|\nu_{2}\rangle&\simeq& \frac{1}{\sqrt{\zeta^2+4A^2}}\left(\zeta|\nu_L\rangle + 2A |S_L^c\rangle \right),\;\;\;\;\;\;\;\;\overline{M_2^L}(K_2\ll K_1)\simeq -\zeta^2K_1 ,\nonumber\\
|\nu_{3}\rangle&\simeq& \frac{1}{\sqrt{\zeta^2+4A^2}}\left(-2A|\nu_L\rangle + \zeta |S_L^c\rangle \right),\;\;\;\;\;\;\overline{M_3^L}(K_2\ll K_1)\simeq -A^2K_1 
\label{HDOstates2}
\end{eqnarray}
where $A\equiv \frac{K_2}{K_1}$ is used. From above one can define the mixing angle between active and sterile neutrinos as $\tan\theta\equiv\frac{2A}{\zeta}$.

Let us consider $\Lambda_1\sim 10^{11}$ GeV, $\Lambda_2\sim 1$ TeV and $m_S\sim 0.01$ eV (which does not affect the masses and the mixing angle much unless $m_S$ is of the order $m_{\nu\nu^c}$). Then $\zeta$ and $A$ are of the order of $10^{-7}$ and $10^{-8}$, respectively. From Eq.~(\ref{HDOstates2}), with the set of values taken for the parameters above, we have $\overline{M_1^H} =10^4$ GeV, and $\nu_1$ and $\nu_L^c$ decouple from the others. The masses for $\nu_2$ and $\nu_3$ are approximately $0.1$ eV and $2\times 10^{-3}$ eV, respectively, with $\tan\theta\simeq 0.19$ ($\theta\simeq 10.6^o$, which is big enough to produce the active-sterile mixing required by the LSND data). Indeed, unlike the masses, the mixing angle is very sensitive to the exact value of $K_2$. The above values are for $K_2=2\times 10^{-3}$ eV. Taking $10^{-3}$ eV instead would render the angle half as large. However, the mass for $\nu_3$  becomes  $2\times 10^{-3}$ eV while leaving $\overline{M_2^L}(K_2\ll K_1)$ unchanged. The main point is that it is possible to have two light states having both active and sterile components with small mixing compatible with the solar, atmospheric and LSND neutrino experiments. 

We comment on the case with $\Lambda_2=0$ as a limiting case of the above discussion. The important feature is that the coupling between $\nu_L^c$ and $S_L^c$ disappears. This renders the above  $\nu_3$ state even lighter\footnote{As a matter of fact, it is massless unless the $M_3^L$ term is included. So, negligible $M_3^L$ with respect to $m_S$ is considered below.} and leaves the other states unchanged. However, the $\nu_2$ and $\nu_3$ states of Eq.~(\ref{HDOstates2}) will be completely different. $\nu_3$ will be almost a pure $S_L^c$ state and decouples when $m_S$ is considered negligible (and taken to be zero). The $\nu_2$ and $\nu_1$ states have a very small mixing, of the order of $\zeta\sim 10^{-7}$, between $\nu_L$ and $\nu_L^c$. For the case where $m_S$ is not negligible, we should take into account $|M_3^L|$, which is  $\frac{2\lambda_1 v_3 m_{ee^c}}{m_{EE^c}}$ in the ALR model and $\frac{2m_{\nu\nu^c}^{\prime} m_{ee^c}^{\prime}}{m_{EE^c}^{\prime}}$ in the Inert case. The only modification will be in $\nu_2$ and $\nu_3$ states and, in order to have masses smaller than 1 eV, $M_3^L$ is allowed to be at most one order of magnitude bigger.\footnote{$m_S$ is assumed to be around 0.01 eV. As before, masses around 1 eV give physical masses also close to 1 eV and  very large mixing.} However, in this case the mixing between active-sterile flavor states would be too small ($\leq 3^{ o}$) while the masses are $\pm 0.1$ eV. Let us consider the case where $M_3^L$ is at least two orders of magnitude smaller than $m_S$.\footnote{Indeed, it doesn't matter how small $M_3^L$ is. The mixing angle is not sensitive to  the $M_3^L$ parameter and is very stable. $M_3^L$ doesn't affect much the masses of $\nu_2$ and $\nu_3$ either. So $S_L^c$ can be safely considered a pseudo-Dirac particle.} For $\Lambda_1\sim 10^{11}$ GeV, the masses of $\nu_2$ and $\nu_3$ are 0.1 eV and $10^{-3}$ eV, respectively with $\tan\theta\simeq 0.1$ $(\theta\sim 5.7^{ o})$. So, this case also yields a framework with two light states with almost fixed active-sterile mixing angle
regardless of how small the Majorana mass of $S_L^c$ is. However, the case with non-zero but small $\Lambda_2$ compared to $\Lambda_1$ yields a mixing very sensitive to the value of $K_2$  mainly due to the existence $K_{12}$ coupling in the matrix. The possibility of having $\Lambda_2=0$ (or, in general, one of vanishing scale) has an advantage over the other cases discussed above as it may not always be possible to have two intermediate scale vev's for both of the singlet fields whose masses could be nonzero.

The discussion in this section can be generalized to three generations in a straightforward manner.  
However, one must be concerned about dangerous flavor changing neutral current interactions.  
Since the Higgs sectors of the models include three sets of Higgs bosons, one for each generation,  
the Glashow-Weinberg theorem \cite{Glashow:1976nt} will be violated leading to tree-level flavor 
changing neutral currents mediated by neutral Higgs bosons.  In addition, lepton universality will 
be broken due to mixing between leptons and the $SU(2)_{R(R^{\prime})}$ gaugino.   One can, of course, 
fine-tune the relevant couplings or make the relevant Higgs fields very heavy, but these solutions 
are unnatural.   There are alternatives discussed in the literature \cite{Ma:1988zq,Umemura:1989wq}.
If one chooses a basis 
such that only one neutral Higgs field, say the third generation field, gets a vev, and one also 
considers a discrete symmetry which distinguishes between generations, then there will be no mediation of 
flavor-changing neutral currents between the first two generations.  This can be achieved by 
assigning even parity for the third family Higgs fields and odd parity for those of the first two 
families.  A classification of such generational symmetries has been done \cite{Umemura:1989wq}.  
Unlike the quark 
sector, it is not possible to remove all flavor changing neutral interactions from the lepton sector 
within the above framework.   Bounds on such interactions involving the tau sector are much 
weaker than those involves muon-electron interactions, and such models may be 
phenomenologically acceptable.   

We note that the discussion in this study can also be carried out within the context of the 
Additional Neutral Fermion (ANF) method.  This would have results similar to the HDO method.   
However, our main point is to show that, in the neutral lepton sectors of the ALR and Inert models, 
it is possible to have a framework in which sterile neutrinos exist naturally having small mixings 
with the active neutrinos, consistent with the LSND result.  As shown above, the HDO method 
allows us to realize such a framework and so it is unnecessary to extend the discussion  to the 
more complicated ANF method as well.

\section{Conclusions}\label{sec:conc}

It is possible that ongoing neutrino experiments, such as MiniBooNe, will 
make the necessity of one or 
more sterile neutrinos unavoidable.   A feature of $E_6$ models is that 
the fundamental representation 
of the group contains a number of isosinglets that would be natural 
candidates for such neutrinos.    It 
is important to analyze these models to see if the various neutral 
fermions can give rise to an acceptable 
phenomenology.  In an earlier paper \cite{Frank:2004vg}, we considered $E_6$ subgroups 
which contain an extra $SU(2)$ 
group, concentrating on the Alternative Left-Right (ALR) and Inert models, 
and we examined the 
neutrino spectrum in a non-supersymmetric framework. 

In that paper \cite{Frank:2004vg}, it was shown that both the ALR and Inert models predict 
neutrino sectors which are phenomenologically 
unacceptable.  The lightest state always contained only isosinglets, and 
each generation contained 
isodoublet neutrino states with masses of the order of the $Q=2/3$ quark 
mass.    Three methods that 
alleviated these problems were discussed.  The first was the Discrete 
Symmetry (DS) method, in which a 
discrete symmetry is imposed to eliminate the tree-level Dirac mass. 
 Dirac neutrino masses can only 
be generated at one-loop, and the parameters can easily be adjusted to 
give masses in the correct mass range.  However, there were still  very 
light isosinglet masses, and no mixing with the isodoublets.   The second 
method, the Higher 
Dimensional Operators (HDO) method, required additional Higgs fields 
and an intermediate scale.   This 
method used dimension-5 operators to remove the very light isosinglets 
and thus the lightest neutrino 
states were isodoublets in the correct mass range.  An interesting feature 
of this model was that the 
coupling of the isodoublet neutrinos to the W-boson is somewhat 
suppressed.  Finally, the Additional 
Neutral Fermion (ANF) method required the existence of new particles as 
well as discrete symmetries 
and was able to accommodate mixing between the light isodoublet 
neutrinos and the sterile neutrinos. 

In this paper, we have considered the supersymmetric version of the ALR 
and Inert models.  An 
attractive feature of the supersymmetrization of the models is that the 
Higgs  fields can be taken to be 
supersymmetric partners of some of the exotic neutral fermions in the {\bf 27}-plet. 
The ALR and Inert 
model symmetries then constrain the allowed terms in the 
superpotential.   If one assumes that R-parity is 
conserved, then one finds the mass matrix divides into two sectors.   In 
the $R=+1$ sector, the active neutrino gets a 
large mass, of the order of the $Q=2/3$ quark mass, and mixes 
maximally 
with the isosinglet $\nu^c_L$.   In the $R=-1$ sector, one finds two heavy 
states
and one very light isosinglet state.  Thus one has the same problems as 
in the non-supersymmetric case.   However, there is an interesting 
alternative.  It is possible to mix the two sectors through soft R-parity 
violating terms.   Several cases were analyzed, and it was found that the 
mass and mixing problems still exist.  The only change is that there can 
now be a small mixing between the active neutrino and one of the 
isosinglet 
neutrinos.

Thus, it was necessary to go beyond the minimal content of these models.  
In the Discrete Symmetry method, a discrete symmetry, which is 
generation-dependent, was used to eliminate the Dirac neutrino mass at 
tree-level.  As in the non-supersymmetric picture, one-loop corrections 
can 
give a mass in the right mass range.  Mixing with the isosinglet 
$\nu^c_L$, however, remains maximal, and there is no substantial mixing 
with other isosinglet neutrinos.   We next considered the Higher 
Dimensional Operators method, in which an intermediate scale is 
introduced, 
as well as some isosinglet Higgs fields from a ${\bf\overline{27}}$ representation 
of $E_6$.   Various cases were considered, and it was shown that a fully 
acceptable model, with masses and mixing angles in the 
phenomenologically 
preferred region, can be obtained.  We also briefly discussed the 
generation of tree-level flavor-changing neutral currents due to the 
proliferation of Higgs doublets in the models, and noted that the 
currents can be eliminated in the quark and $(e-\mu)$ sectors, but not 
entirely in the $\tau$ sector.

\begin{acknowledgments}
M.F. and I.T. thank NSERC of Canada for support under grant number 0105354, and  M.S. thanks the National Science Foundation 
under grant PHY-0243400.
\end{acknowledgments}

\appendix*\section{}
\label{sec:Appendix}

In this Appendix, we give the masses and the corresponding states for the case in which $\Rslash$-parity is included in the ALR model. We give the description for ALR but the results apply for Inert as well. The results are summarized in Table \ref{Rviolation}. We would like to comment on the assumptions used to get these results. Recall that $M_3^L\equiv 2\lambda_1 v_3 m_{ee^c}/m_{EE^c}$.
\begin{table}[htb]
	\caption{The eigenstates and masses through $\Rslash$-parity for different hierarchies among $m_S, M_3^L$, and $m_{\nu\nu^c}$. Here $\xi$ and $\Theta$ are defined as  $\frac{m_S}{m_{\nu\nu^c}}$ and $\frac{M_3^L}{6 m_{\nu\nu^c}}\left(2+\frac{\xi^2-2}{\sqrt{1+\xi^2}}\right)$, respectively.} \label{Rviolation}     
\begin{center}
    \begin{tabular}{cccc}
       \hline\hline
\vspace{0.1cm}
\hspace*{-1cm}$\rm Cases $ &  & \hspace*{-2.5cm}$\rm States$ &\hspace*{-2.5cm} $\rm Masses$  \\ 
\hline 
                               &  Light   &  \multicolumn{1}{l}{$|\nu_3\rangle \simeq |S_L^c\rangle$}     & \multicolumn{1}{l}{$M_3^L\simeq \frac{2\lambda_1 v_3 m_{ee^c}}{m_{EE^c}}$ }       \\
$m_S\ll |M_3^L|\ll m_{\nu\nu^c}$ \\ 
                               &  Heavy    &  \multicolumn{1}{l}{$|\nu_{1,2}\rangle\simeq \frac{1}{\sqrt{2}}\left(|\nu_L\rangle \pm |\nu_L^c\rangle \right)$}  &  \multicolumn{1}{l}{$\overline{M_{1,2}^H}\simeq \pm m_{\nu\nu^c}$}       \\
\\
\hline 
                               &  Light     &  \multicolumn{1}{l}{$|\nu_3\rangle \simeq |S_L^c\rangle-\xi |\nu_L^c\rangle$}    & \multicolumn{1}{l}{$\overline{M_3^L}\simeq\frac{2\lambda_1 v_3 m_{ee^c}}{m_{EE^c}}\left(1-\frac{2}{3}\xi^2\right)$}      \\
$|M_3^L|\ll m_S\ll m_{\nu\nu^c}$  \\
                               &  Heavy     &  \multicolumn{1}{l}{$|\nu_{1,2}\rangle\simeq \frac{\left(|\nu_L\rangle \pm |\nu_L^c\rangle \pm \xi |S_L^c\rangle \right)}{\sqrt{2}}$}     &  \multicolumn{1}{l}{$\overline{M_{1,2}^H}\simeq \pm m_{\nu\nu^c}\left(1+\frac{1}{2}\xi^2\right)$}        \\
\\
\hline
                                & Light  &  \multicolumn{1}{l}{$|\nu_3\rangle \simeq \frac{-\xi |\nu_L^c\rangle +|S_L^c\rangle}{\sqrt{1+\xi^2}}$}  &  \multicolumn{1}{l}{$\overline{M_3^L}\simeq \frac{2\lambda_1 v_3 m_{ee^c}}{3m_{EE^c}}(1+\frac{2-\xi^2}{\sqrt{1+\xi^2}})$}     \\
$|M_3^L|\ll m_S\sim m_{\nu\nu^c}$  \\
                                & Heavy     &  \multicolumn{1}{l}{$|\nu_{1,2}\rangle\simeq \frac{\left( \sqrt{1+\xi^2}|\nu_L\rangle \pm |\nu_L^c\rangle + \xi |S_L^c\rangle \right)}{\sqrt{2(1+\xi^2)}}$}  &   \multicolumn{1}{l}{$\overline{M_{1,2}^H}\simeq  m_{\nu\nu^c}\sqrt{1+\xi^2}\left( \pm 1 + \Theta\right)$} \\
\\
\hline \hline
        \end{tabular}
        \end{center}
\vskip -0.5cm
\end{table}

For the case~(i), $m_S\ll |M_3^L|\ll m_{\nu\nu^c}$, we keep only $O(M_3^L/m_{\nu\nu^c})$ and $O(m_S/M_3^L)$ terms but not terms $O(m_S/m_{\nu\nu^c})$ and $O(m_S/M_3^L)$. The next order correction to the masses and the states listed in Table \ref{Rviolation} are of the order $O((m_S/M_3^L)^2)$ and $O(m_S/m_{\nu\nu^c})$ which are presumed very small and negligible. Due to the absence of $O(m_S/M_3^L)$ terms in mass eigenvalues, $\nu_{1,2}$ does not have a $S_L^c$ component.  

For the case~(ii), $|M_3^L|\ll m_S\ll m_{\nu\nu^c}$, only $O(m_S/m_{\nu\nu^c})$ and $O(M_3^L/m_S)$ terms are kept, such that the orthogonality of $|\nu_1\rangle$ and $|\nu_2\rangle$ is satisfied up to the order of $(\frac{m_S}{m_{\nu\nu^c}})^2$. If $m_{\nu\nu^c}$ is allowed to have values less than eV, the physical neutrino states $\nu_{1,2}$ in the second row of Table~\ref{Rviolation} would give a ${\bf 3+1}$ structure. This requires extreme fine-tuning. Furthermore, we should note that the sterile state $|\nu_L^c\rangle + \frac{m_S}{m_{\nu\nu^c}} |S_L^c\rangle$  mixes maximally with $|\nu_L\rangle$, which would be inconsistent with the constraints from the LSND result. In this case $S_L^c$ has a small mixing with $\nu_L^c$. The lightest state $\nu_3$ could have the desired light mass, but it is totally sterile and has no chance to be detected. 

Finally, the case~(iii), $|M_3^L|\ll m_S\sim m_{\nu\nu^c}$, which  is possible when we consider fairly large soft-term couplings $\mu_A$, is obviously a modification of the case~(ii) when $m_{\nu\nu^c}$ and $m_S$ are comparable. To get the results given in the third row of Table~\ref{Rviolation}, we neglected terms of orders $O(|M_3^L|/m_{\nu\nu^c})$ and $O(|M_3^L|/m_S)$. The sterile state is now composed of $\nu_L^c$ and $S_L^c$ mixing almost maximally  and the almost purely sterile state indeed has an active component whose mixing is proportional to $O(|M_3^L|/m_{\nu\nu^c})$. Their masses are also modified accordingly.


\end{document}